\documentclass[%
 aip, 
 amsmath,amssymb,
 reprint,%
]{revtex4-1}

\usepackage{graphicx}
\usepackage{dcolumn}
\usepackage{bm}

\usepackage[utf8]{inputenc}
\usepackage[T1]{fontenc}
\usepackage{mathptmx}
\usepackage{etoolbox}
\usepackage{hyperref}
\hypersetup{
    colorlinks=true,
    linkcolor=black,
    citecolor=blue,
    urlcolor=blue,
}

\makeatletter
\def\@email#1#2{%
 \endgroup
 \patchcmd{\titleblock@produce}
  {\frontmatter@RRAPformat}
  {\frontmatter@RRAPformat{\produce@RRAP{*#1\href{mailto:#2}{#2}}}\frontmatter@RRAPformat}
  {}{}
}%
\makeatother
\begin{document}

\preprint{AIP/123-QED}

\title{\color{blue}Strain Engineering of Epitaxial Pt/Fe Spintronic Terahertz Emitter}
\author{Rahul Gupta}
\email{rahul.gupta@angstrom.uu.se; rahulguptaphy@gmail.com}
\affiliation{Department of Materials Science and Engineering, Uppsala University, Box 35, SE-751 03 Uppsala, Sweden}

\author{Ebrahim Bagherikorani}%
\affiliation{Center of Excellence for Applied Electromagnetic Systems, University of Tehran, Iran}%

\author{Venkatesh Mottamchetty}
\affiliation{Department of Materials Science and Engineering, Uppsala University, Box 35, SE-751 03 Uppsala, Sweden}%

\author{Martin Pavelka}
\affiliation{Department of Physics and Astronomy, Box 516, SE-75120 Uppsala, Sweden}

\author{Kasturie Jatkar}
\affiliation{Department of Physics, Stockholm University, SE-10691 Stockholm, Sweden}

\author{Dragos Dancila}
\affiliation{Department of Electrical Engineering, Uppsala University, Box 65, SE-751 03 Uppsala, Sweden}

\author{Karim Mohammadpour-Aghdam}
\affiliation{Center of Excellence for Applied Electromagnetic Systems, University of Tehran, Iran}%

\author{Anders Rydberg}
\affiliation{Department of Materials Science and Engineering, Uppsala University, Box 35, SE-751 03 Uppsala, Sweden}
\affiliation{Department of Physics and Astronomy, Box 516, SE-75120 Uppsala, Sweden}

\author{Rimantas Brucas}
\affiliation{Department of Materials Science and Engineering, Uppsala University, Box 35, SE-751 03 Uppsala, Sweden}
\affiliation{Ångström Microstructure Laboratory, Uppsala University, Box 35, SE-751 03 Uppsala, Sweden}

\author{Hermann A. Dürr}
\affiliation{Department of Physics and Astronomy, Box 516, SE-75120 Uppsala, Sweden}

\author{Peter Svedlindh}%
\affiliation{Department of Materials Science and Engineering, Uppsala University, Box 35, SE-751 03 Uppsala, Sweden}

\date{\today}

\begin{abstract}
Spin-based terahertz (THz) emitters, utilising the inverse spin Hall effect, are ultra-modern sources for the generation of THz electromagnetic radiation. To make a powerful emitter having large THz amplitude and bandwidth, fundamental understanding in terms of microscopic models is essential. This study reveals important factors to engineer the THz emission amplitude and bandwidth in epitaxial Pt/Fe emitters grown on MgO and MgAl$_2$O$_4$ (MAO) substrates, where the choice of substrate plays an important role. The THz amplitude and bandwidth is affected by the induced strain in the Fe spin source layer. On the one hand, the THz electric field amplitude is found to be larger when Pt/Fe is grown on MgO even though the crystalline quality of the Fe film is superior when grown on MAO. This is because of the larger defect density, smaller  electron relaxation time and lower electrical conductivity in the THz regime when Fe is grown on MgO. On the other hand, the bandwidth is found to be larger for Pt/Fe grown on MAO and is explained by the uncoupled/coupled Lorentz oscillator models. This study provides an insightful pathway to further engineer metallic spintronic THz emitters in terms of proper choice of substrate and microscopic properties of the emitter layers. 
\end{abstract}

\maketitle

\section{Introduction}
The terahertz (THz) radiation lies in the frequency range from 0.1 to 30 THz and can be generated by nonlinear crystals \cite{yang1971generation,vediyappan2019evaluation}, photoconductive antennas \cite{smith1988subpicosecond,venkatesh2014optical,isgandarov2021intense}, air plasma \cite{cook2000intense,PhysRevLett.97.103903,bagley2018laser}, etc. However, all these emitters are costly, relatively difficult to make and contain phonon absorption gaps in the THz frequency domain. Broadband THz sources find applications in imaging\cite{valuvsis2021roadmap}, communication \cite{federici2010review} and applied science\cite{cheon2019detection}. Recently, it has been demonstrated that THz radiation can also be generated by metallic ferromagnetic (FM)/non-ferromagnetic (NFM) bilayer thin films, known as spin-based THz emitters (STEs) \cite{seifert2016efficient,gupta2021co2feal}, where the FM and NFM layers are used as spin source and spin sink, respectively. The advantages of STEs are low cost, ease of fabrication and that the emitted THz radiation extends up to 40 THz without any spectral absorption in the THz frequency domain.

The integration of the spin degree of freedom in electronic devices is the key ingredient in spintronics devices. For future, to increase the writing speed of spin based random access memories and the storage density in highly packed spin based memory architectures, antiferromagnetic systems have recently been proposed as they do not generate magnetic stray fields \cite{shi2020electrical}. However, to understand the fundamental physics \cite{jin2015accessing} and spin dynamics of such systems, a broadband THz source is required without any spectral absorption in the THz frequency domain. For example, the second order spin dynamics referred to as spin nutation \cite{thonig2017magnetic,neeraj2020inertial,PhysRevB.103.104404} and optically induced spin-orbit torque both lie in the THz frequency domain \cite{PhysRevResearch.3.023116}.
\begin{figure*}
\centering
    \includegraphics[width=18cm]{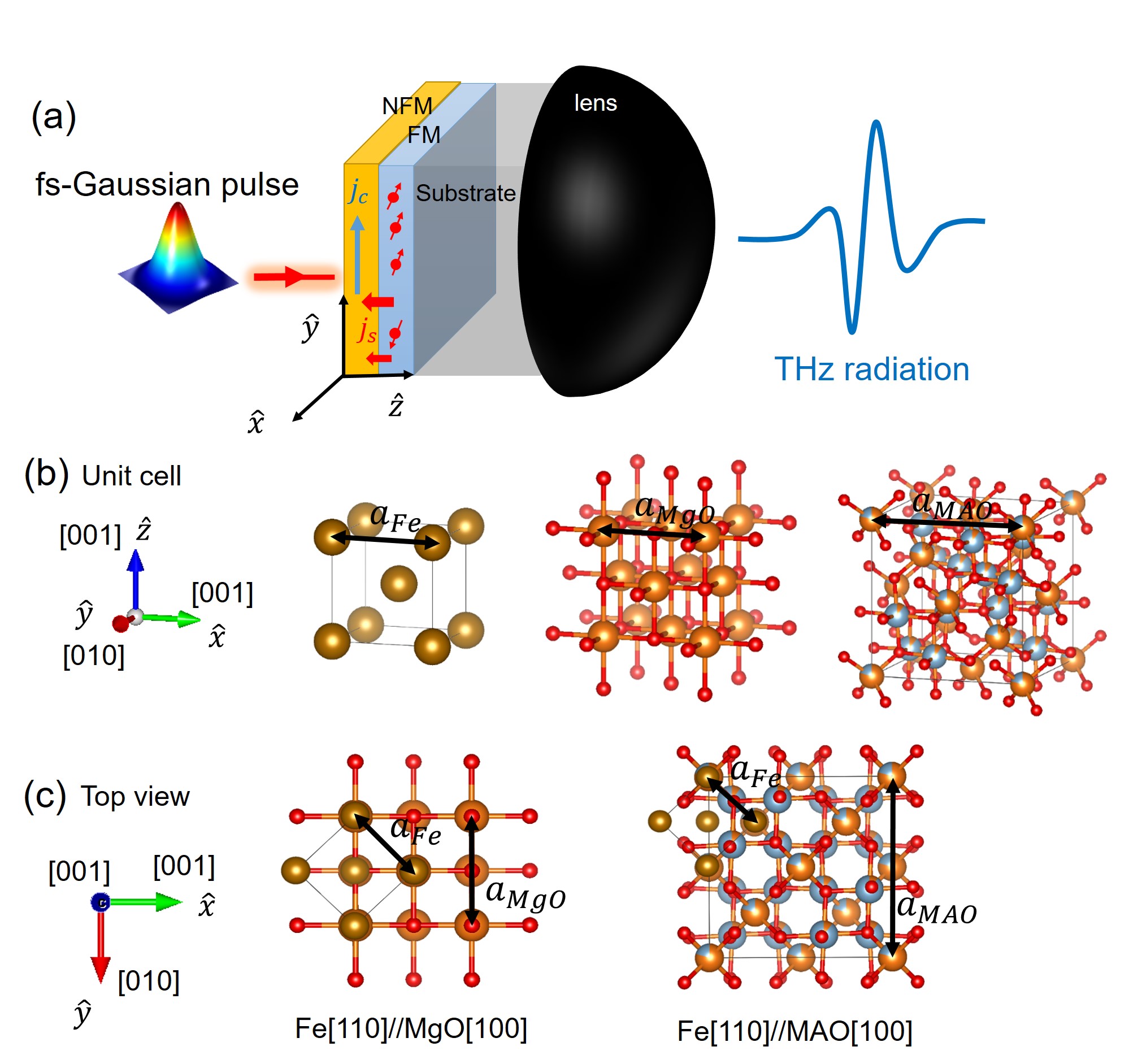}
    \caption{(a) Schematic of NFM/FM spin-based terahertz emitter on substrate with hemispherical lens, where the laser is incident from the NFM side, (b) unit cell of Fe, MgO and MAO, and (c) top view of Fe on MgO and MAO.}
    \label{fig:my_label}
\end{figure*}

For THz generation in STEs, a femtosecond (fs) laser pulse creates an optically induced spin current in the FM layer, which superdiffuses into the NFM layer via the bilayer interface \cite{PhysRevLett.105.027203,choi2015thermal,choi2014spin}. This spin current is converted into a charge current pulse via the inverse spin Hall effect (iSHE) assisted by a strong spin-orbit coupling (SOC) in the NFM layer. Furthermore, the charge current pulse generates THz electromagnetic dipole radiation due to the acceleration/deceleration of charge carriers in the NFM layer, as indicated in Fig. (1a). 

The THz electric field amplitude ($E_{THz}$) is described by Ohm’s law as follows,
\begin{equation}
    E_{THz} =  j_c \cdot Z_{eff}(\omega).
\end{equation}
The first term $j_c$ is the charge current, which depends on the spin to charge conversion efficiency ($\theta$) and the optically induced spin current ($j_s$) and is defined as,
\begin{equation}
    j_c=-e \int_{0}^d \theta(z) j_s(z,\omega)dz.
\end{equation}

The second term in Eq. (1) is the effective impedance of the STE, which depends on the choice of substrate and the conductivity of the STE and is defined as,
\begin{equation}
    \frac{1}{Z_{eff}(\omega)} = \int_{0}^d \sigma_{STE}(z,\omega)dz +  \frac{1}{Z(\omega)},
\end{equation}
where $\sigma_{STE}$ and $d$ are the conductivity in the THz frequency regime and  the total thickness of the metallic STE layers, respectively. $Z(\omega)$ is the impedance with contributions from the substrate and hemispherical lens. The hemispherical lens is used to collect and collimate the emitted THz radiation from the source. From Eqs. (1-3), it is clear that the emitted THz radiation depends on several factors, such as generation of spin current in the FM layer, spin to charge conversion efficiency of the NFM layer, conductivity of the FM and NFM layers, and the impedance of the substrate. 
Previously, it has been demonstrated that the THz emission from STEs depends on the spin to charge conversion efficiency of the NFM layer \cite{seifert2016efficient}. Therefore, to keep this factor constant in our study, we choose Pt as spin sink layer, which is a large SOC material. Alternatively, two dimensional-transition metal dichalcogenides (2D-TMDs) can be used as a spin sink layer instead of Pt because of larger SOC strength \cite{husain2020emergence}. However, there are several challenges to grow 2D-TMDs on large scale substrates. Thus, we have chosen Pt as spin sink layer together with Fe as spin source layer in our STEs. The choice of Fe as source layer and choice of substrates are explained in the forthcoming section. 

According to the theory of impedance matching, the refractive index of the hemispherical lens and substrate should be same to avoid the losses at the substrate/lens interface when the fs laser is incident from the NFM side as shown in Fig. (1a). Si and Al${_2}$O${_3}$ hemispherical lenses\cite{van1989high} are commonly used to collimate the THz beam. Therefore, MgO and Al${_2}$O${_3}$ substrates come into the picture in order minimize the impedance mismatch as both substrates have a refractive index similar to that of Si and Al${_2}$O${_3}$ lenses. However, Al${_2}$O${_3}$ single crystal substrates have a hexagonal lattice, which implies that there is a challenge to grow Fe epitaxially on Al${_2}$O${_3}$ substrates. Moreover, Nenno \textit{et al.} showed that the THz emission is larger in epitaxial Fe than in polycrystalline Fe based STEs, which were grown on MgO and Al$_2$O$_3$ substrates, respectively \cite{nenno2019modification}. However, the conductivity of an epitaxial STE as compared to a polycrystalline ditto is expected to be larger due to less scattering centers. As per the thickness dependent THz emission model described in Ref. \cite{torosyan2018optimized,gupta2021co2feal}, it is expected that the epitaxial STE should exhibit lower THz emission as compared to the polycrystalline STE.
Thus, to investigate the conductivity effect of epitaxial Fe and Pt, we choose MgO and MgAl$_2$O$_4$ (MAO) as substrates together with an Si hemispherical lens in our THz emission set-up. So far, the scientific challenges of STEs have been addressed in various ways; optimization of STE thickness \cite{torosyan2018optimized}, different combinations of spin source and spin sink materials\cite{seifert2016efficient}, ferro\cite{wu2017high}-, ferri- and antiferromagnetic\cite{zhou2019orientation,qiu2021ultrafast} systems as spin source, and different laser pulse excitation wavelength\cite{herapath2019impact}, but not in terms of conductivity and crystal quality of the STEs. 

In this article, we focus our attention on the crystal quality of the spin source layer. The crystal quality depends on choice of substrate for the epitaxial growth of different layers,  where MgO and MAO substrates have been chosen for the growth of the spin source (Fe) and the spin sink (Pt) layers.

\begin{figure}
\centering
    \includegraphics[width=10cm]{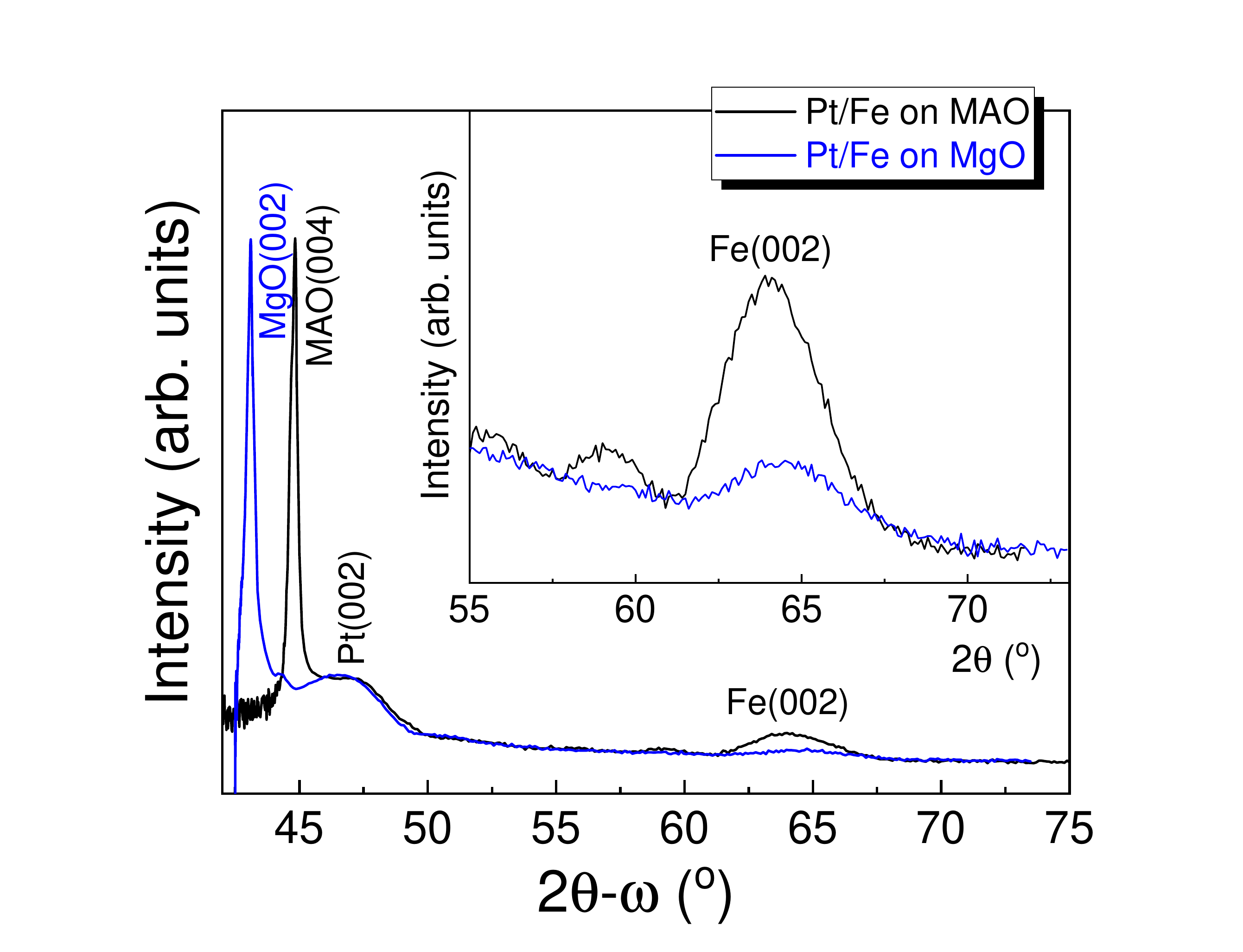}
    \caption{X-ray diffraction pattern for Pt(3)/Fe(4) on MgO and MAO substrates in the perpendicular diffraction vector geometry.}
    \label{fig:my_label}
\end{figure}

\begin{figure}
\centering
    \includegraphics[width=10cm]{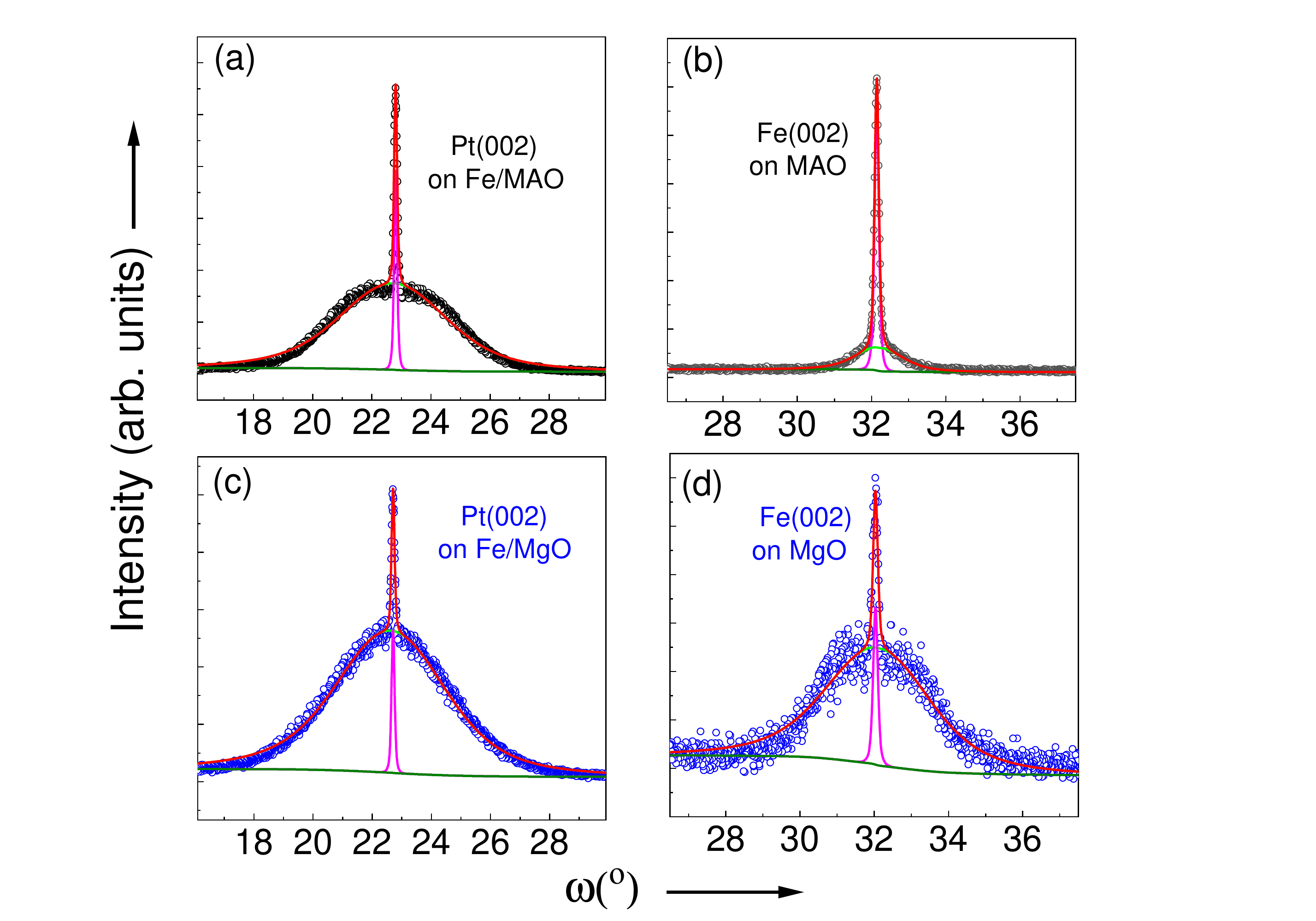}
    \caption{Rocking curves measured along (002) Bragg plane for (a) Pt(3) on Fe(4)/MAO, (b) Fe(4) on MAO, (c) Pt(3) on Fe(4)/MgO, and (d) Fe(4) on MgO. Black and blue correspond to experimental data. Green, magenta, red, and olive correspond to Lorentzian fitted, Gaussian fitted, a mixture of Lorentzian and Gaussian fitted, and background fitted data, respectively.}
    \label{fig:my_label}
\end{figure}

\begin{figure*}
    \centering
    \includegraphics[width=18cm]{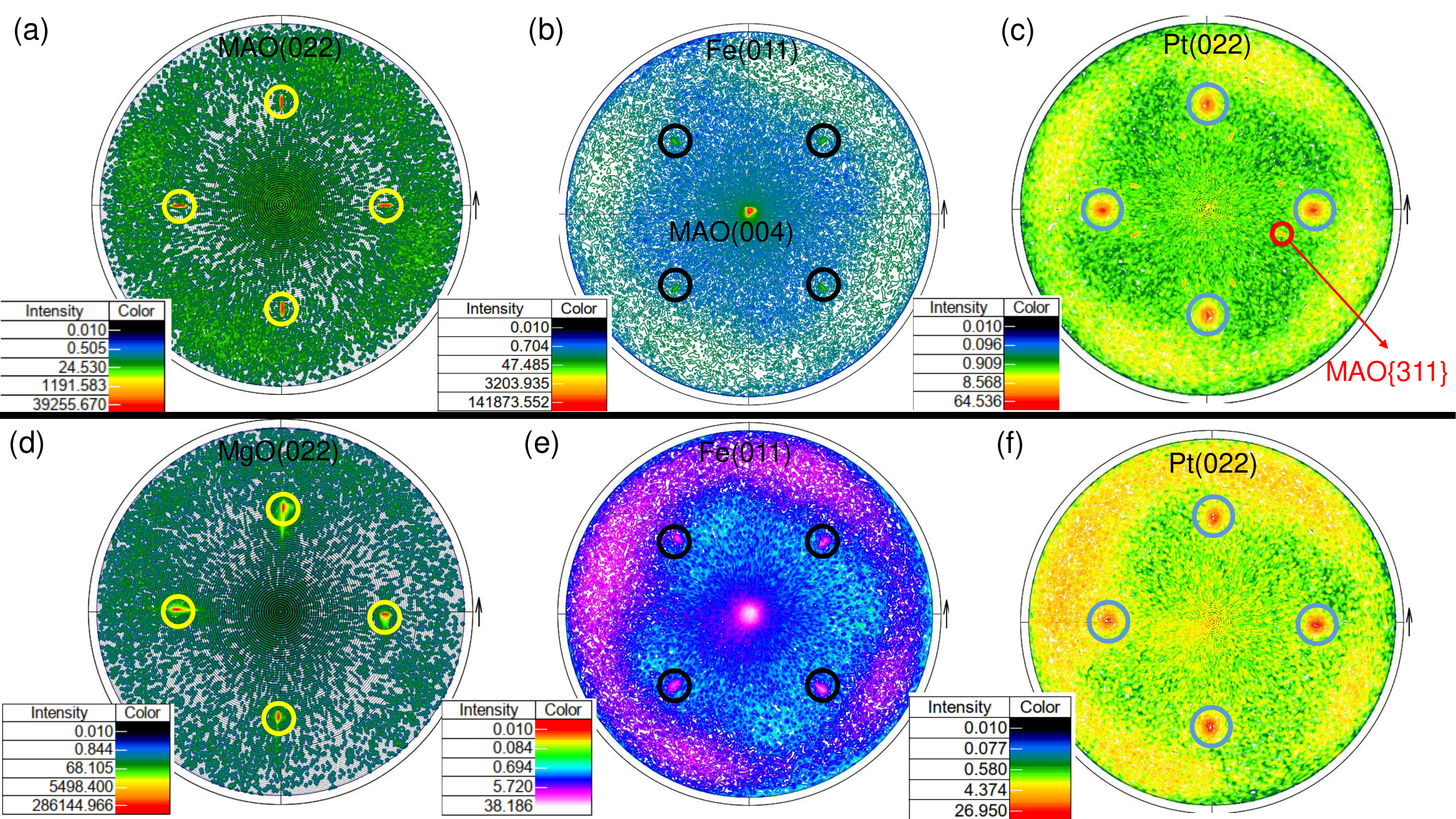}
    \caption{Pole figures for Pt(3)/Fe(4)/MAO (top) and Pt(3)/Fe(4)/MgO (bottom). Pole figures along (a) MAO(022), (b) Fe(011), (c) Pt(022), (d) MgO(022) (e) Fe (011), and (f) Pt(022) Bragg planes.}
    \label{fig:my_label}
\end{figure*}

\begin{figure}
    \centering
    \includegraphics[width=10cm]{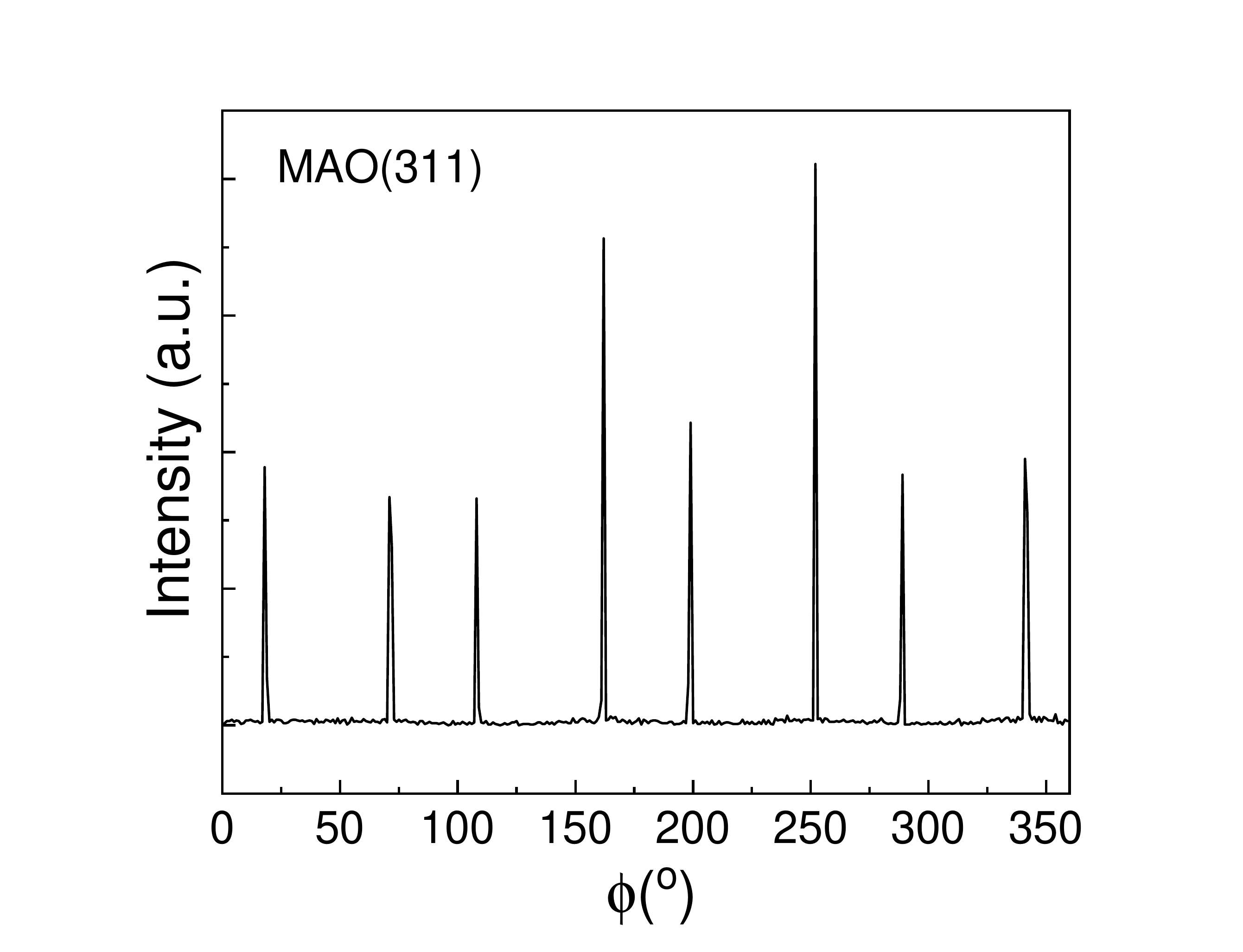}
    \caption{$\phi$-scan along MAO(311) Bragg plane for Pt(3)/Fe(4)/MAO.}
    \label{fig:my_label}
\end{figure}


\begin{figure*}
\centering
    \includegraphics[width=18cm]{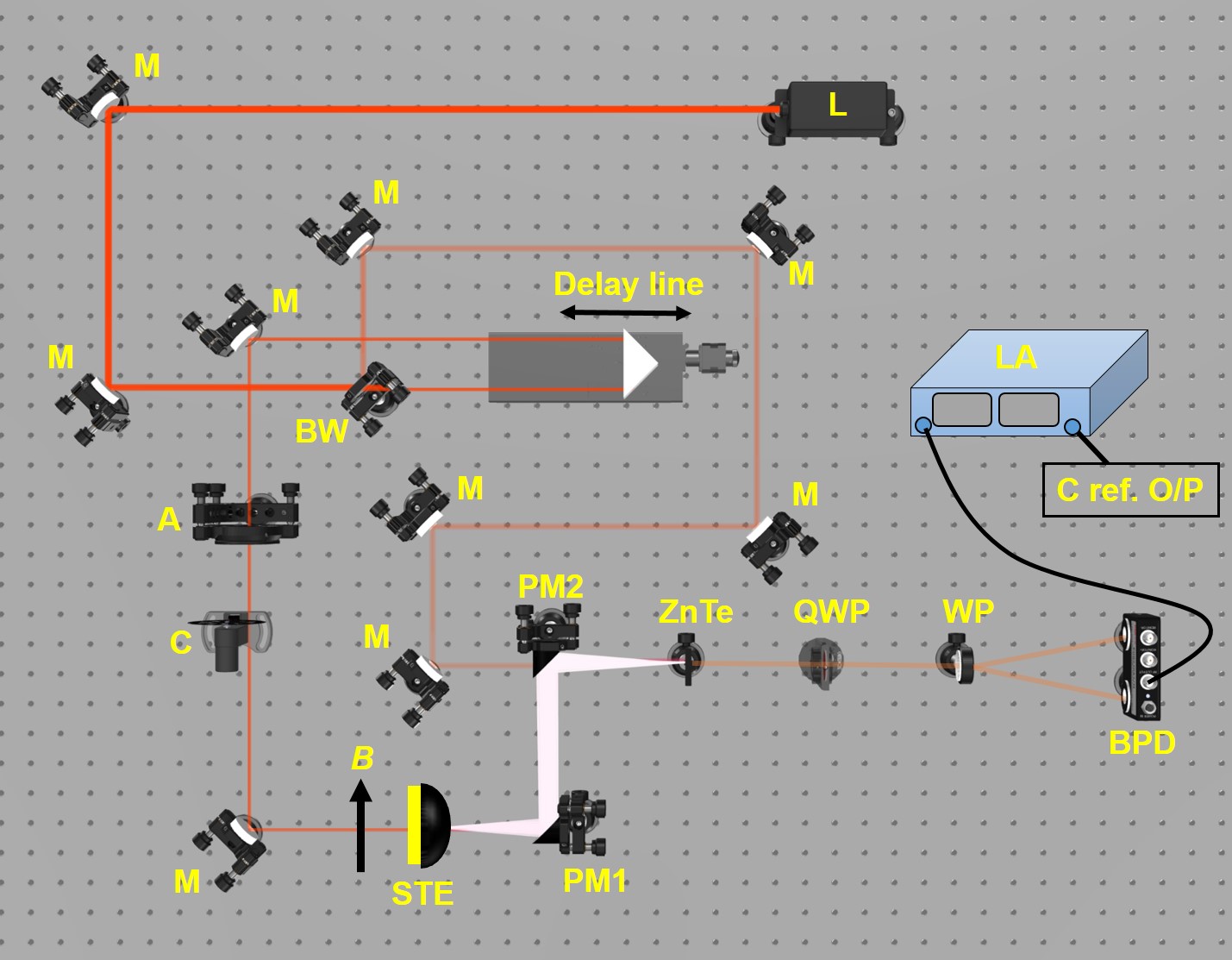}
    \caption{Schematic of THz-TDS set up using electro optical detection. L: fs laser (55 fs, 825 mW, 800 nm, 80 MHz), M: mirror, BW: Brewster window, A: attenuator, C: chopper, B: in-plane applied magnetic field, STE: Pt/Fe spin-based terahertz emitter, PM1 and PM2: parabolic mirrors, ZnTe: zinc telluride crystal, QWP: quarter wave plate, WP: Wollaston prism, BPD: balanced photo diodes, LA: lock-in amplifier, C ref. O/P: chopper reference output.}
    \label{fig:my_label}
\end{figure*}

\begin{figure*}
\centering
    \includegraphics[width=18cm]{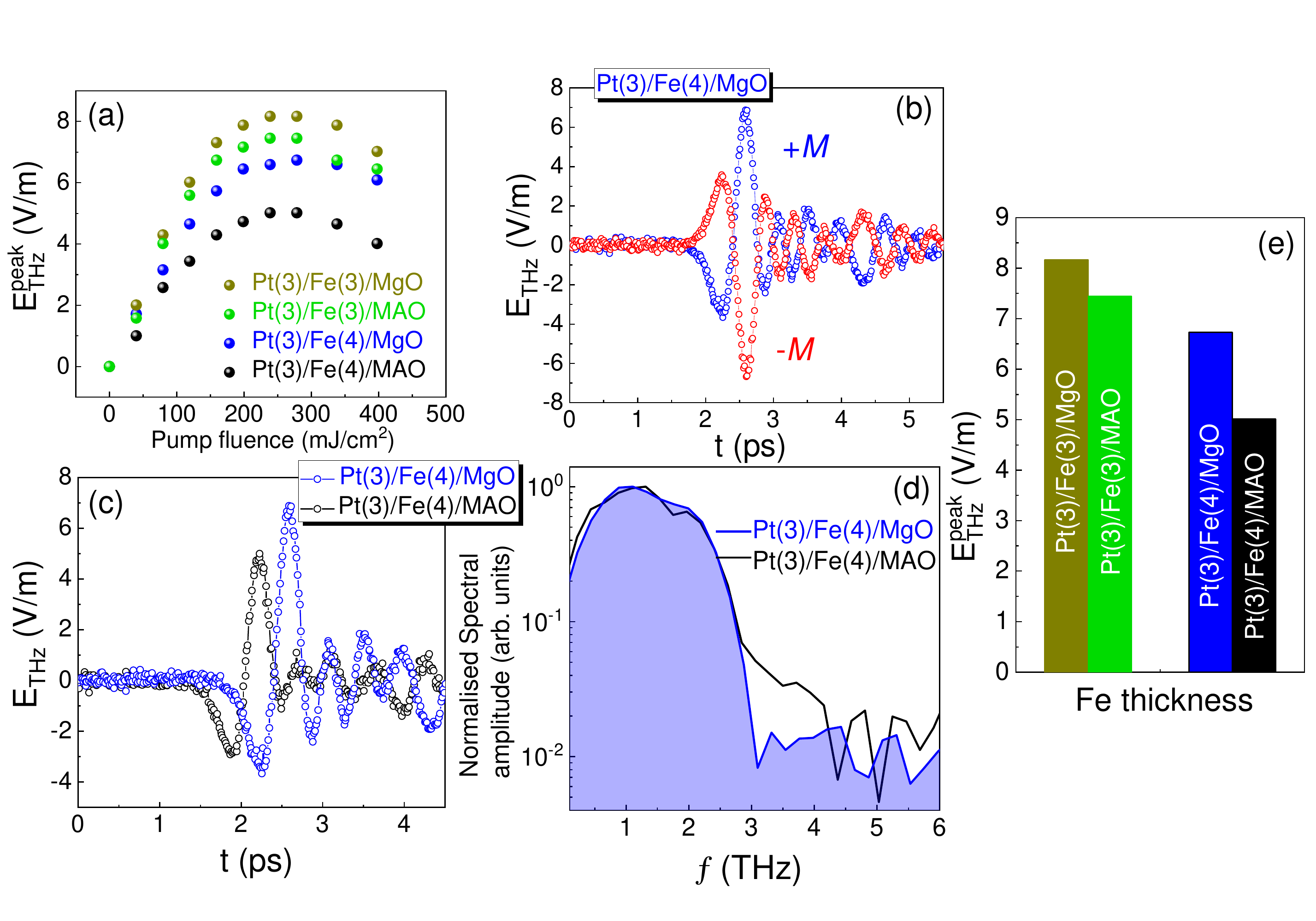}
    \caption{(a) THz electric field  peak amplitude (E$^{peak}_{THz}$) as function of pump fluence, (b) THz electric field  peak amplitude for positive and negative applied magnetic field for Pt(3)/Fe(4) STE on MgO substrate, (c) THz electric field  peak amplitude for Pt(3)/Fe(4) STE on MgO substrate (blue) and on MAO substrate (black). The time domain signal is shifted for better visualization, (d) fast Fourier transform of time domain signal for Pt(3)/Fe(4) STE on MgO substrate (blue) and on MAO substrate (black), and (e) comparison of THz electric field  peak amplitude at 278 mJ/cm$^2$ fluence for different Pt/Fe STEs on MgO and MAO substrates.}
    \label{fig:my_label}
\end{figure*}

\section{Model system and experimental details}

Body centred cubic (bcc) Fe is known to grow epitaxially on MgO single crystal substrates. The in-plane lattice constants ($a$) of MgO and Fe are 4.21 Å and 2.86 Å, respectively. This enables epitaxial growth of Fe with a 45$^\circ$ in-plane rotation on the MgO substrate as the distance between Fe atoms along the [110] direction is 4.04 Å, which provides a lattice mismatch [$\in$ = (a$_{substrate}$-a$_{film}$)/a$_{substrate}$] between Fe and MgO of $\sim$4\%.

MAO exhibits a spinel crystal structure with an in-plane lattice constant of 8.1 Å. In analogy with MgO, bcc Fe is expected to grow epitaxially with a 45$^\circ$ in-plane rotation with respect to the MAO substrate in order to minimise its lattice mismatch ($\sim$0.4\%). The unit cells of Fe, MgO, and MAO and top view of Fe on MgO and MAO are shown in Fig. (1b-c).

Pt/Fe bilayer thin films were grown on MgO(001) and MAO(001) single crystal substrates using a CMS-18 Kurt J. Lesker Company magnetron sputtering system at a base pressure lower than 5$\times$10$^{-8}$ Torr. The substrates (0.5 mm thick) were ultrasonically cleaned in acetone and isopropanol, introduced into the growth chamber, outgassed at 600$^\circ$C for 1 hour under vacuum conditions, and finally cooled down to deposition temperature (300$^\circ$C). Then, Fe(3, 4 nm) films were deposited on MgO and on MAO substrates by DC sputtering from an Fe target (purity of 99.95\%)  under an argon pressure of 5 mTorr, at a deposition rate of 0.3 Å/s. Later, Pt(3 nm) layers were deposited on Fe/MgO and Fe/MAO using a Pt target (purity of 99.99\%) under the same argon pressure, at a deposition rate of 0.6 Å/s. Deposition rates were precisely calibrated using X-ray reflectivity (XRR) measurements together with  simulations using the GenX3 software\cite{bjorck2007genx}.


X-ray diffraction measurements in specular mode with Cu K$_{\alpha1}$ radiation ($\lambda$=1.54 Å) were used to check the crystal quality of the Fe and Pt layers. In 2$\theta$-$\omega$ scans as shown in Fig. (2), only Bragg reflections from Fe(002) and Pt(002) planes are found for the MgO and MAO substrates, providing an indication of the epitaxial nature or high degree of crystal orientation for both layers. Rocking curves for the Pt and Fe films on both MgO and MAO substrates are  shown in Fig. (3). There are two features appearing in the rocking curves of the Pt(002) and Fe(002) Bragg planes; a sharp peak (Gaussian contribution) and a broad hump (Lorentzian contribution). The sharp peaks in all rocking curves originate from the substrate and the broad humps are due to the Fe and Pt thin films. This observation shows that the ﬁlm close to the substrate/ﬁlm interface has a higher crystal quality than the subsequently grown ﬁlm. This agrees with previously demonstrated feature\cite{lindahl2009epitaxial}. To extract the full width at half maximum (FWHM) values, a mixture of Gaussian and Lorentzian functions was used to fit the experimental data. The FWHM values for the Pt(002) and Fe(002) Bragg planes are listed in Table (I). On the one hand, the FWHM for Fe grown on MAO is found to be  $\sim$2 times smaller than for Fe grown on MgO, implying less strain in Fe films grown on MAO as compared to Fe grown on MgO. This result is also in accordance with the expected lattice mismatch for the two substrates as mentioned earlier. On the other hand, the FWHM for the Pt(002) Bragg plane is found to be similar for Fe/MgO and Fe/MAO. 

\begin{table}[]
    \centering
    \begin{tabular}{c|c|c}
     & FWHM ($^{\circ}$) & FWHM ($^{\circ}$) \\
    Bragg planes & Lorentzian contribution & Gaussian contribution \\
    \hline
    \hline
    Pt(002) on Fe/MgO & 4.396 (22) & 0.122 (03)\\
    Pt(002) on Fe/MAO & 4.316 (44) & 0.123 (02)\\
    Fe(002) on MgO & 3.328 (43) & 0.149 (07)\\
    Fe(002) on MAO & 1.527 (21) & 0.136 (01)\\
    \end{tabular}
    \caption{Full width at half maximum for Pt and Fe films along (002) Bragg plane without accounting for the constant instrumental resolution.}
    \label{tab:my_label}
\end{table}

A perfect epitaxial film with cubic structure is expected to have a four-fold symmetry. Therefore, to further confirm the epitaxial nature of the Pt and Fe layers, we have used pole figure measurements using point focus primary optics for both Pt/Fe/MAO and Pt/Fe/MgO. The results shown in Fig. (4) clearly evidence the four-fold symmetry for the substrates as well as for the different layers, which confirms the epitaxial nature of the films and the single crystal nature of the substrates. Here, it is noticed that the four-fold symmetry of Fe has a 45$^\circ$ in-plane rotation with respect to the four-fold symmetry of the substrates as shown in Fig. (4a-b, 4d-e). Similarly, the four-fold symmetry of Pt has also a 45$^\circ$ in-plane rotation with respect to the four-fold symmetry of Fe as shown in Fig. (4b-c, 4e-f). However, an eight-fold symmetry can be seen in the Pt pole figure when grown on Fe/MAO, indicated by the red circle in Fig. (4c). This eight-fold symmetry corresponds to the  MAO(311) Bragg plane as shown in Fig. (5). To conclude, Pt and Fe grow epitaxially on both MgO and MAO substrates. However, the epitaxial quality of Fe is superior when grown on MAO. In the next section, we will show how strain induced effects in the Fe layer affect the THz emission in Pt/Fe STEs. 

THz time domain spectroscopy (THz-TDS)  has been used to measure the THz emission from the Pt/Fe STEs. The schematic of the experimental set-up used for THz generation and detection is shown in Fig. (6). The employed laser source (Ti: sapphire) delivers 55 fs pulses with energy 9 nJ at a repetition rate of 80 MHz. A zinc telluride (ZnTe) nonlinear crystal (Egorov Scientific ZnTe-10-10-1-AR) is used as THz source and detector to calibrate the THz-TDS set-up. The THz emission from the Pt/Fe STE on both MgO and MAO substrates has been measured by replacing the ZnTe source with the Pt/Fe STEs. The laser beam is incident on a Brewster window (BW) and the transmitted part (pump)  is focused on the Pt/Fe STE, while the reflected part is guided to the ZnTe detector (probe). The employed pump and probe beam powers were $\sim$0-100 mW and $\sim$100 mW for emission and detection of THz radiation, respectively. Upon illumination of the  Pt/Fe STE, it emits THz radiation in the presence of an in-plane  magnetic field (85 mT or 0.67$\times$10$^5$ A m$^{-1}$). The diverging THz pulses obtained from the STEs is collected, collimated and focused onto the ZnTe [(110) orientation, 1 mm thickness] detector by a Si hemispherical lens and parabolic mirrors (PM1 and PM2).  The probe beam from the BW is focused on to the ZnTe detector by a plano convex lens of focal length 20 cm and the emitted THz radiation is detected by electro-optic sampling using a quarter wave plate (QWP), a Wollaston prism (WP) and a pair of balanced photo diodes (BPD). The BPD’s  analog output is connected to the voltage input of the lock-in amplifier which is using a reference frequency of 6 kHz from the mechanical chopper. The temporal profile of the THz radiation is measured by varying the delay of the probe beam with respect to the THz pulse.

\section{Results and Discussion}
The THz emission of Pt(3)/Fe(4)/MgO, Pt(3)/Fe(4)/MAO, Pt(3)/Fe(3)/MgO, and Pt(3)/Fe(3)/MAO has been measured as a function of pump fluence; the results are shown in Fig. (7a). Here it is noticed that the THz emission is found to be larger for Pt/Fe  grown  on MgO as compared to MAO. As shown in Fig. (7b), the polarity of the THz emission from Pt(3)/Fe(4)/MgO is inverted while reversing the magnetic field (magnetization) direction, which confirms that the THz emission from the Pt/Fe STEs originates from the iSHE. A comparison of the THz emission of Pt(3)/Fe(4) on MgO and MAO substrates in time domain and the corresponding frequency domain spectra are shown in Fig. (7c-d). Here, on the one hand, the THz emission amplitude is found to be larger for Pt(3)/Fe(4) deposited on MgO, while the THz bandwidth is found to be larger for Pt(3)/Fe(4) deposited on MAO (0.2-3 THz for Pt(3)/Fe(4)/MgO and 0.2-4 THz for Pt(3)/Fe(4)/MAO). The bandwidth is limited by the comparably large pulse duration ($\sim$55 fs) of the incident pump. 
The same behavior in terms of bandwidth and THz emission amplitude has been observed by replacing the ZnTe detector with low temperature GaAs detector\cite{conferpaper}. To understand the differences in THz amplitude and bandwidth, we focus our attention to the microscopic properties of STE layers such as the defect density, the electrical conductivity in the THz frequency regime, the effect on spin memory loss (SML) at the Pt/Fe interface, the refractive index and the transmittance of the MgO and MAO substrates in the THz regime, which will be discussed in the forthcoming sections.


\begin{table*}
    \centering
    \begin{tabular}{c|c|c|c|c}
   Spintronic & E$^{peak}_{THz}$ & $\sigma_{DC}$ (MS/m) & $\sigma_{DC}$ (MS/m) & $\tau$ (fs)\\
    THz emitter & (V/m) & from vdP method
    & from Drude fit & from Drude fit\\
    \hline
    \hline
    Pt(3)/Fe(3)/MAO & 7.44 & 3.31276 $\pm$	0.00138 & 2.60 $\pm$ 0.93 & 41.06 $\pm$ 1.1\\
    Pt(3)/Fe(3)/MgO & 8.16 & 2.08784 $\pm$	0.00065 & 1.95 $\pm$ 0.78 & 13.47 $\pm$ 2.5\\
    Pt(3)/Fe(4)/MAO & 5.01 & 3.22115 $\pm$	0.00114 & 2.76 $\pm$ 0.68 & 24.3 $\pm$ 1.2\\
    Pt(3)/Fe(4)/MgO & 6.73 & 2.49569 $\pm$	0.00017 & 2.31 $\pm$ 0.73 & 9.92 $\pm$ 3.5\\
    \end{tabular}
    \caption{Emitted THz electric field peak amplitude (E$^{peak}_{THz}$), measured DC conductivity $\sigma_{DC}$ and $\sigma_{DC}$ estimated from Drude model fit of AC conductivity data, together with extracted electron scattering time $\tau$ for all four samples.}
    \label{tab:my_label}
\end{table*}

Temperature dependent DC conductivity measurements were used to further examine the strain related effects of the Pt/Fe/MAO and Pt/Fe/MgO STEs. The DC conductivity as a function of temperature was  measured using the van der Pauw method (vdP); the results are shown in Fig. (8). There are two contributions to the conductivity in a metallic layer; electron-phonon scattering and electron-defect scattering \cite{bardeen1940electrical}. At room temperature, both contributions play important roles, while at sufficiently low temperature, the electron-phonon contribution becomes negligible. The constant level of the DC conductivity in Fig. (8) provides qualitative information about the abundance of defects in the material due to electron-defect scattering. Here, it is found that the DC conductivity at low temperature (10 K) is larger for Pt/Fe/MAO, corresponding to a lower defect density in Pt/Fe when deposited on MAO as compared to when deposited on MgO. This difference in defect density is caused by the one order of magnitude difference in lattice mismatch.

\begin{figure}
\centering
    \includegraphics[width=10cm]{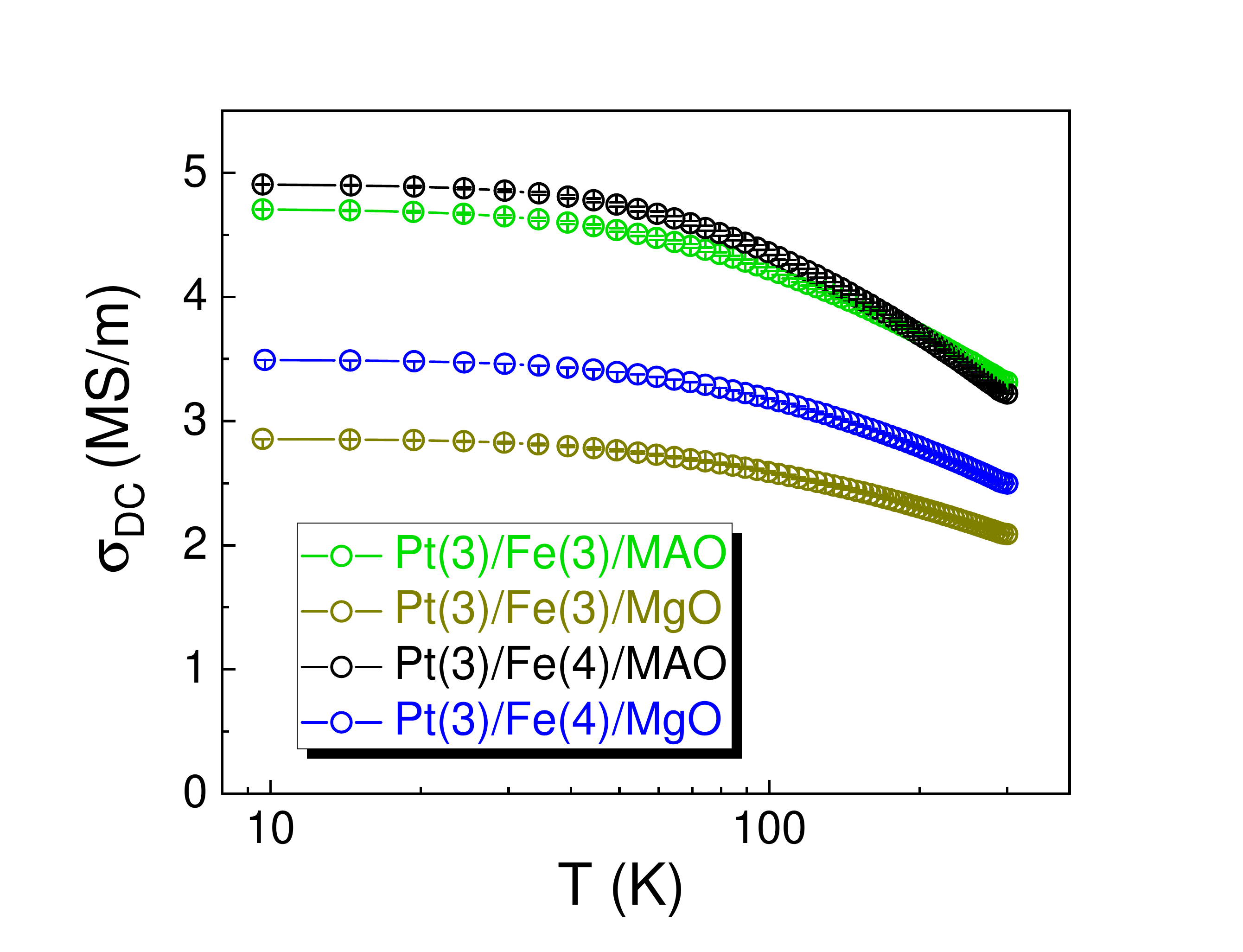}
    \caption{DC conductivity as a function of temperature for Pt/Fe STEs on MgO and MAO substrates.}
    \label{fig:my_label}
\end{figure}

\begin{figure}
    \centering
    \includegraphics[width=8cm]{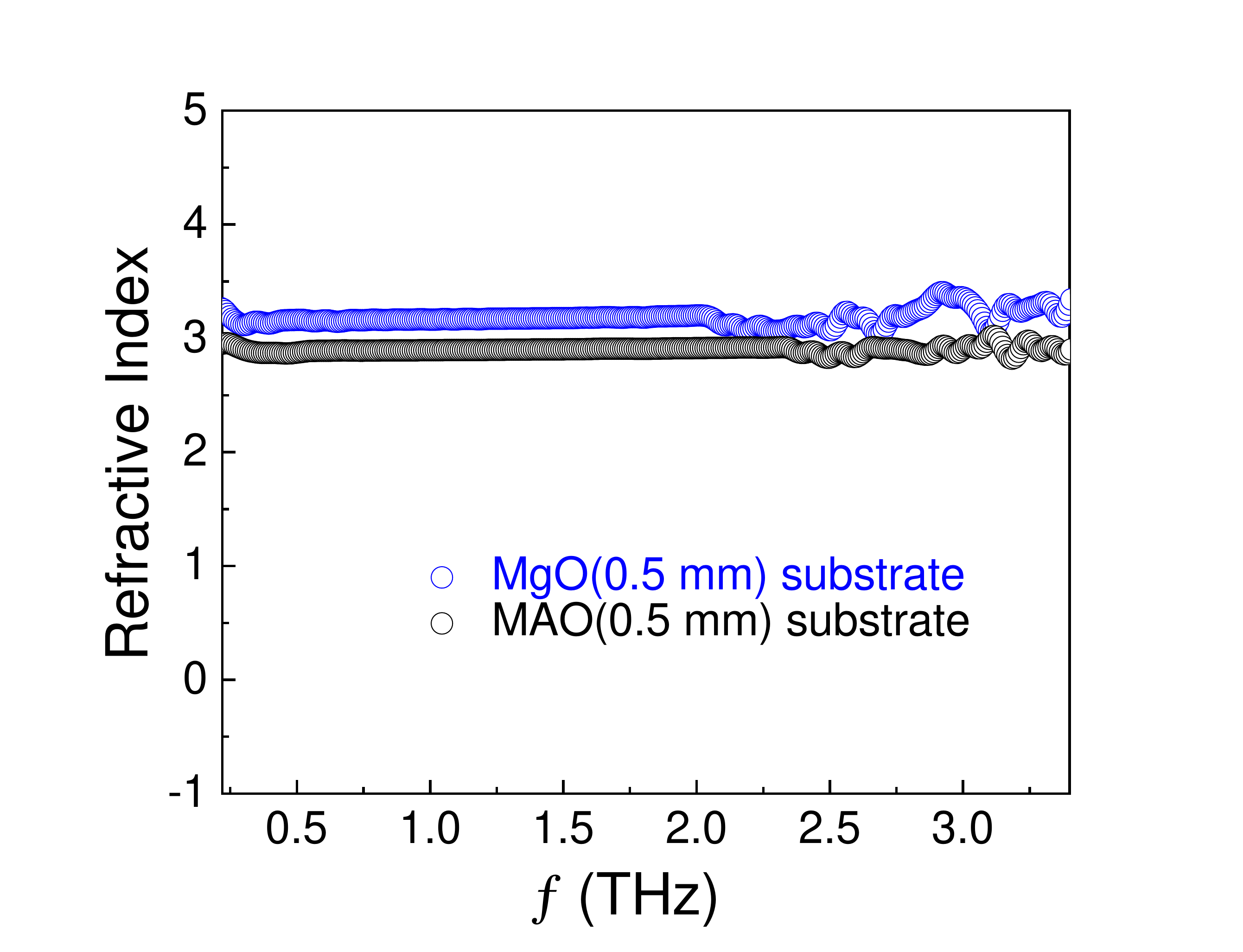}
    \caption{Observed refractive index of MgO and MAO substrates in THz frequency regime, respectively.}
    \label{fig:my_label}
\end{figure}

From Eqs. (1-3) it is clear that the emission of THz radiation depends on the electrical conductivity of the STE layers in the THz frequency regime. Therefore, we further inverstigated electrical properties of the STEs by  exploiting THz-TDS. For this purpose, a TeraFlash Pro system manufactured by TOPTICA GmbH \cite{TOpica} was used to measure THz transmission through STEs.
First, the THz electric field $E_{substrate}(t)$ transmitted through the substrate without the STE layers was recorded as reference, followed by recording the transmitted electric field pulse for the substrate with the STE layers, referred to as $E_{STE/substrate}(t)$. In these experiments, one must account for multiple reflections of the THz pulse within the substrate. For this purpose, one records the time dependence of the electric field which shows echoes in the time range of 10 ps. Such echoes are removed in the analysis before applying the fast Fourier transform (FFT). The ratio of the FFTs of $E_{substrate}(t)$ and $E_{STE/substrate}(t)$ gives the complex transmittance \(\tilde T(\omega)\) of the STE layers. The transmittance is related to the complex conductivity [$\sigma (\omega)$] of the thin film via Tinkham's formula \cite{PhysRev.108.243},
\begin{equation}
    \tilde T(\omega)=\frac{FFT[E_{STE/substrate}(t)]}{FFT[E_{substrate}(t)]}=\frac{1+n_{substrate}}{1+n_{substrate}+Z_{0}\cdot d \cdot {\sigma_{AC} }(\omega)} \,
    \label{eq:Thinkham}
\end{equation}    

where $n_{substrate}$ is the refractive index of the substrate, \(Z_{0}=377\,\Omega\) is the free space impedance, $d$ and  $\sigma_{AC}(\omega)$ are the total thickness and the complex conductivity of the STE layers, respectively. The thickness $d$ has been extracted using XRR, which is discussed in the forthcoming section. The refractive index $n_{substrate}$ has been measured using THz-TDS spectroscopy and the method to extract the $n_{substrate}$ is mentioned in the supplementary information of Ref. \cite{gupta2021co2feal}. The real part of the refractive index of the MgO and MAO substrates are found to be 3.17 and 2.89, respectively as shown in Fig. (9), which implies similar impedance mismatch between the substrates and the Si lens. These parameters are used as input in Eq. (4) to extract  $\sigma_{AC}(\omega)$. The extracted real and imaginary parts of the AC conductivity are shown in Fig. (10). Such measurements have been previously used in AC conductivity characterization of various types of materials such as crystalline and amorphous metallic thin films\cite{PhysRevLett.117.087205}. The behavior of the complex electrical conductivity of metals as a function of frequency can be described by the Drude model \cite{drude1900electrontheory,ashcroft1976solid},

\begin{equation}
\sigma_{AC}(\omega)=\frac{\sigma_{DC}}{1-i \omega \tau}
\label{eq:DM}
\end{equation}
where $\sigma_{DC}$ corresponds to the DC electrical conductivity and $\tau$ is the electron relaxation time. These parameters have been extracted by self-consistent fitting of the real and imaginary parts of $\sigma_{AC}(\omega)$ using Eq. (5).  The Drude model shows a good fit for the AC conductivity obtained from THz-TDS for the Pt/Fe films on MgO and MAO substrates as shown in Fig. (10). Table (II) summarizes the values of $\sigma_{DC}$ from the vdP method, $\sigma_{DC}$ and $\tau$ from the Drude fit of the AC conductivity data. These result are consistent with thickness dependent measurement of $\sigma_{dc}$ and $\tau$ in Fe thin films \cite{krewer2020thickness}. The DC conductivity values obtained from the AC conductivity data show close matching with those obtained from the vdP method. Here, the real part of the AC conductivity is found to be larger for Pt/Fe on MAO due to smaller defect density as shown in Fig. (10). The electron relaxation time $\tau$ is also found to be larger for Pt/Fe on MAO, implying less electron scattering in the metallic layers and less THz emission amplitude.
\begin{figure}
\centering
    \includegraphics[width=10cm]{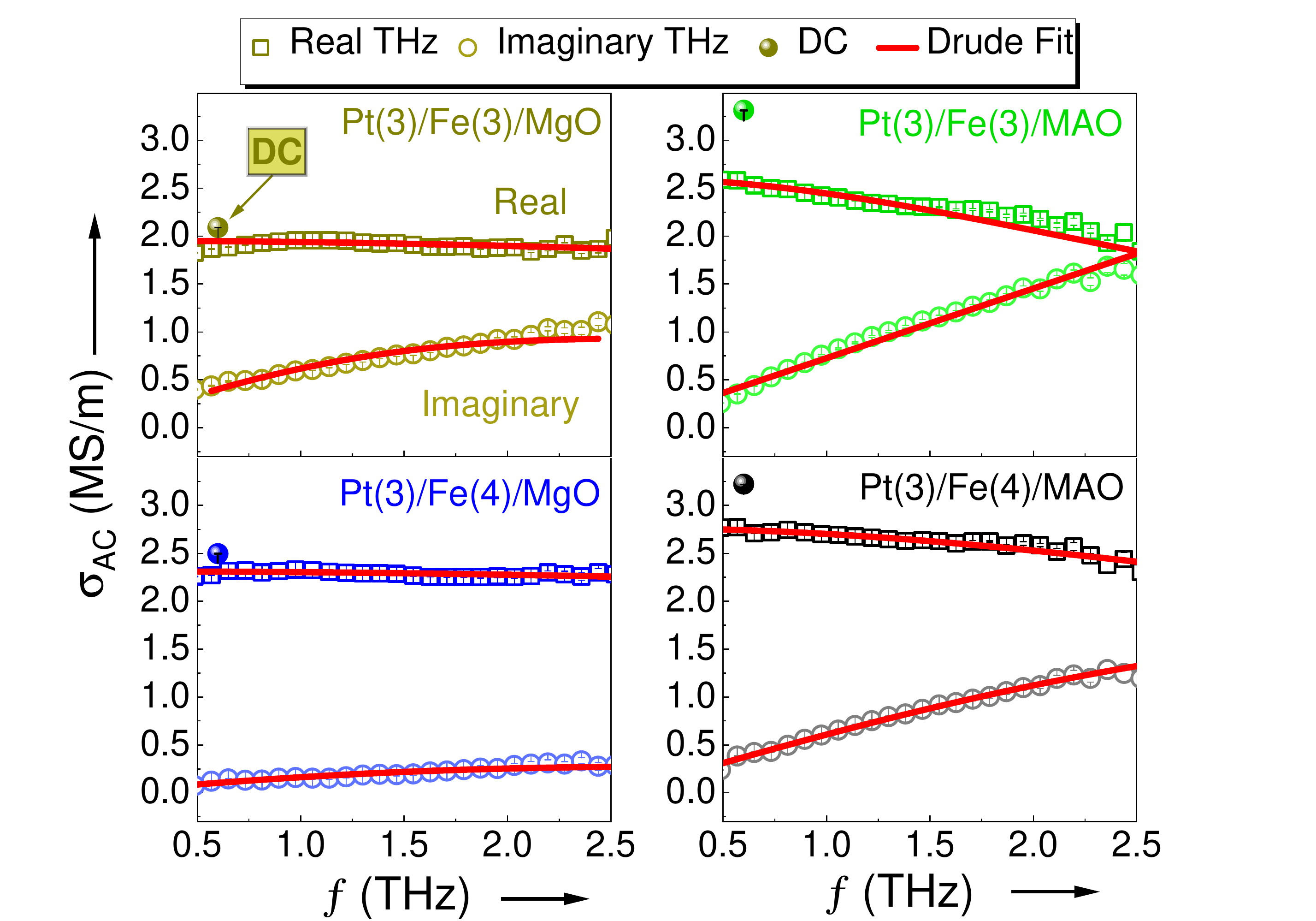}
    \caption{AC conductivity versus frequency in THz regime at room temperature. Empty symbols and lines correspond to observed data and  Drude fitted real and imaginary parts of the AC conductivity, respectively. Solid symbols correspond to measured DC electrical conductivity using the vdP method. The solid symbols are shifted to 0.6 THz for better visualization. }
    \label{fig:my_label}
\end{figure}
However, a difference in THz amplitude for Pt/Fe STEs deposited on MgO and MAO substrates could in principle be explained by a difference in Pt/Fe interface roughness because of SML at the interface. Larger interface roughness results in smaller THz electric field amplitude as there will be less spin current transmitted from Fe to Pt \cite{PhysRevMaterials.3.084415,PhysRevLett.124.087702}. The XRR measurements using the GenX3 software to simulate the experimental data were used to investigate a possible influence of SML; the results are shown in Fig. (11). The interface roughness and thickness of the  individual layers are listed in Table (III). The results show that the Pt/Fe interface roughness is similar for both substrates, which implies the similar SML for both types of Pt/Fe STEs. 
Hence, it can be concluded that the larger THz emission for Pt/Fe deposited on MgO is only explained by the smaller electrical conductivity (electron relaxation time) in THz frequency regime caused by the larger strain in Fe when grown on MgO.

\begin{table}[]
    \centering
    \begin{tabular}{c|c|c|c|c}
Samples &    Layers & Density (FU/Å$^3$) & Thickness (Å) & Roughness (Å) \\
\hline
\hline
 &    Pt(3) & 0.061 (01)  & 33.51 (65)  & 1.97 (08) \\
Pt(3)/Fe(4)/MgO &    Fe(4)     & 0.088 (06)  & 31.22 (67) & 5.02 (80) \\
  &    MgO     & 0.053   & - & 2.07 (40) \\
 \hline
 &    Pt(3)     & 0.064  (01) & 33.65 (40) & 1.70  (07)\\
Pt(3)/Fe(4)/MAO &    Fe(4)     & 0.092  (05) & 30.91  (50) & 4.71  (40) \\
   &    MAO     & 0.015   & - & 1.31 (26) \\
 \hline
   &    Pt(3)     & 0.060  (01) & 34.55 (21) & 2.02  (62)\\
 Pt(3)/Fe(3)/MgO &    Fe(3)     & 0.094  (03) & 22.36  (22) & 4.00  (18) \\
   &    MgO     & 0.053   & - & 2.09 (24) \\
 \hline
 &    Pt(3) & 0.070 (05)  & 33.35 (43)  & 1.64 (03) \\
 Pt(3)/Fe(3)/MAO  &    Fe(3)     & 0.1022 (05)  & 23.08 (38) & 4.37 (32)\\
  &    MAO     & 0.015   & - & 1.19 (11) \\
 \hline
 \hline
    \end{tabular}
    \caption{Parameters obtained for Pt/Fe STEs from the X-ray reflectivity measurements.}
    \label{tab:my_label}
\end{table}

\begin{figure}
    \centering
    \includegraphics[width=10cm]{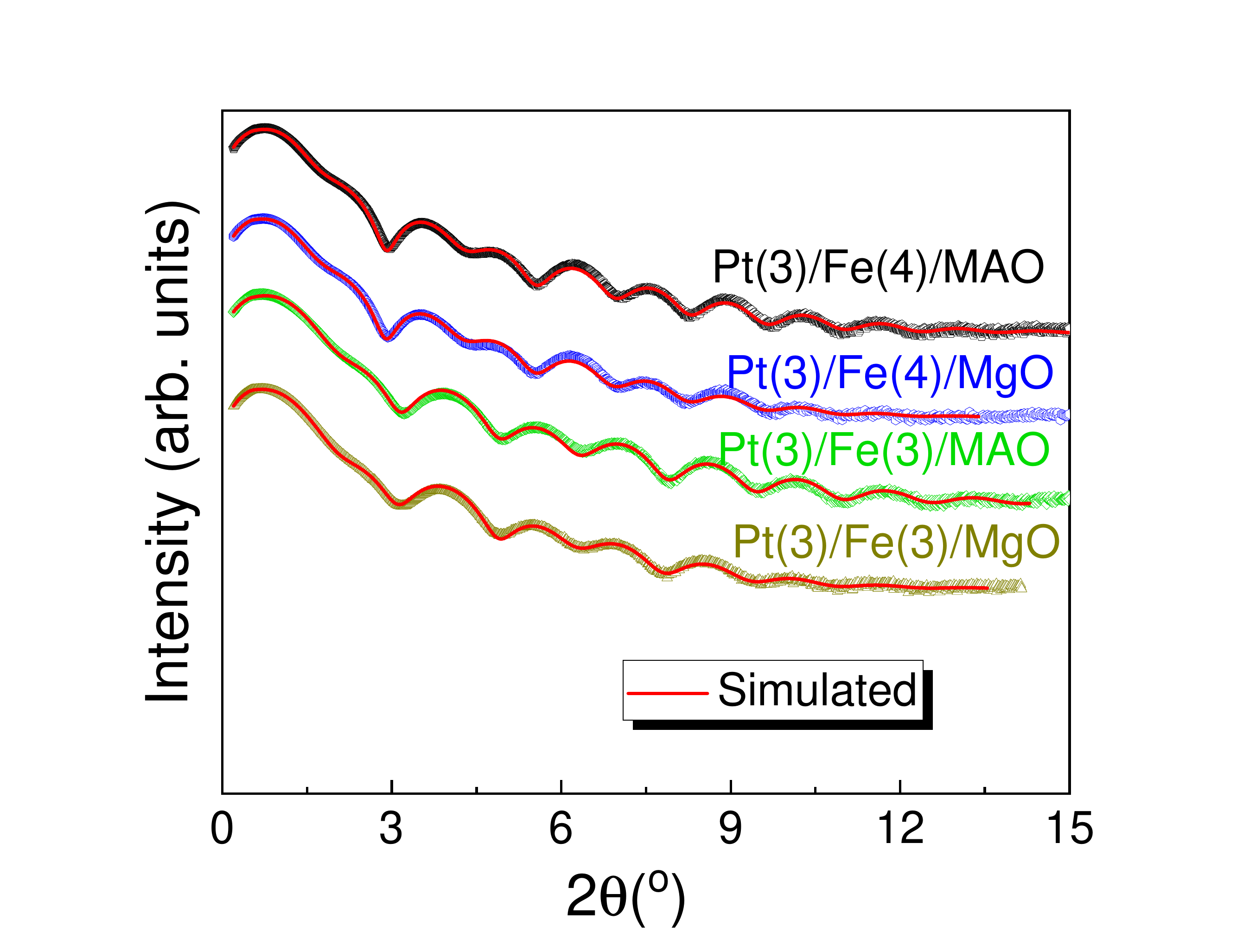}
    \caption{X-ray reflectivity results for Pt/Fe STEs on MgO and MAO substrates. Symbols and lines correspond to observed and simulated data, respectively.}
    \label{fig:my_label}
\end{figure}


To explain the difference in bandwidth between Pt/Fe on MgO and MAO, THz-TDS can be used to measure the THz transmission of the substrates. The THz electric field amplitude in time domain has been measured with and without the substrate placed in the path between the emitter and the detector. The FFT of the time-domain THz signal provides a measure of the complex electric field in the frequency domain and the ratio of the complex electric fields yields the transmission of THz radiation through the substrate and is defined as,

\begin{equation}
    \tilde T_{substrate} = \frac{FFT[E_{substrate}(t)]}{FFT[E_{air}(t)]}
\end{equation}
where $E_{substrate}(t)$ and $E_{air}(t)$ are the complex electric fields in time domain in the presence and absence of the substrate, respectively. The measured transmission in THz regime of the MgO and MAO substrates is shown by black colour in Fig. (12). Here, it is found that both substrates exhibit an oscillatory behavior of the transmittance with more than 70\% transmission of THz radiation. The oscillation appears because of multiple reflections of THz radiation in the substrates. Moreover, it is clearly evident that there is an absorption peak at 3 THz (100 cm$^{-1}$) for MgO, which corresponds to  phonon absorption\cite{komandin2009multiphonon}.

The THz transmittance can be simulated by extracting the permittivity contribution as a function of frequency for the MgO and MAO substrates. The permittivity can be described by the Lorentz oscillator model (LOM), which involves modeling an electron as a driven damped harmonic oscillator. The first order LOM is defined as,
\begin{equation}
    \epsilon_r = \epsilon_\infty + \frac{\Delta \epsilon \nu_{res}^2}{(\nu_{res}^2-\nu^2)+i \gamma \nu}
\end{equation}
where $\epsilon_\infty$ is the high-frequency permittivity, $\Delta \epsilon$ is the permittivity contribution of the first mode, $\nu_{res}$ is the resonance frequency and $\gamma$ is the damping coefficient. According to the simulations, at room temperature, the first order LOM model is sufficient to describe the transmittance spectra for MAO in the 0.3-4 THz frequency range and for MgO in the 0.3-2.5 THz frequency range. The doubly degenerate transverse optical (TO) phonon frequency [$\nu(TO)$] is found to be much larger than the frequency region of interest. Thus, no absorption is found in the simulations within the 0.3-4 THz frequency regime for MAO and within 0.3-2.5 THz for MgO. The parameters for the TO mode for MAO and MgO  are summarized in Table (IV), where $\epsilon_\infty$ = 0 for MAO and 2.95 for MgO. The damping coefficient for the TO mode is found to be $\sim$2 times larger in MAO than in MgO, which explains the larger interaction between Mg, Al and O atoms in MAO than between Mg and O atoms in MgO. A comparison of the simulated and measured THz transmittance for MAO and MgO is shown in Fig. (12).
The simulated THz transmittance is matching with the measured THz transmittance up to 2.5 THz, while a clear difference is observed in the 2.5-4 THz range for MgO. To explain this behavior the coupled LOM is required, which is defined as \cite{komandin2009multiphonon},
\begin{equation}
    \epsilon_r = \frac{s_1(\nu_{2}^2-\nu^2+i\gamma_2 \nu)+s_2(\nu_{1}^2-\nu^2+i\gamma_1 \nu)-2\sqrt{s_1 s_2}(\alpha+i\nu \delta)}{(\nu_{2}^2-\nu^2+i\gamma_2 \nu)(\nu_{1}^2-\nu^2+i\gamma_1 \nu)-(\alpha+i\nu \delta)^2}
\end{equation}

where $s_i$, $\nu_i$ and $\gamma_i$ correspond to strength, resonance frequency and damping coefficient of oscillator $i$, respectively.  $\alpha$ and $\delta$ correspond to the real and imaginary parts of the coupling constant of the two oscillators, respectively. 
The parameters used in the simulations are summarised in Table (IV). Here, one can see that the resonance frequency corresponding to the MgO $D_{11}$ oscillator is found to be 97 cm$^{-1}$, which is very close to previously reported values \cite{komandin2009multiphonon}.

\begin{figure}
    \centering
    \includegraphics[width=8cm]{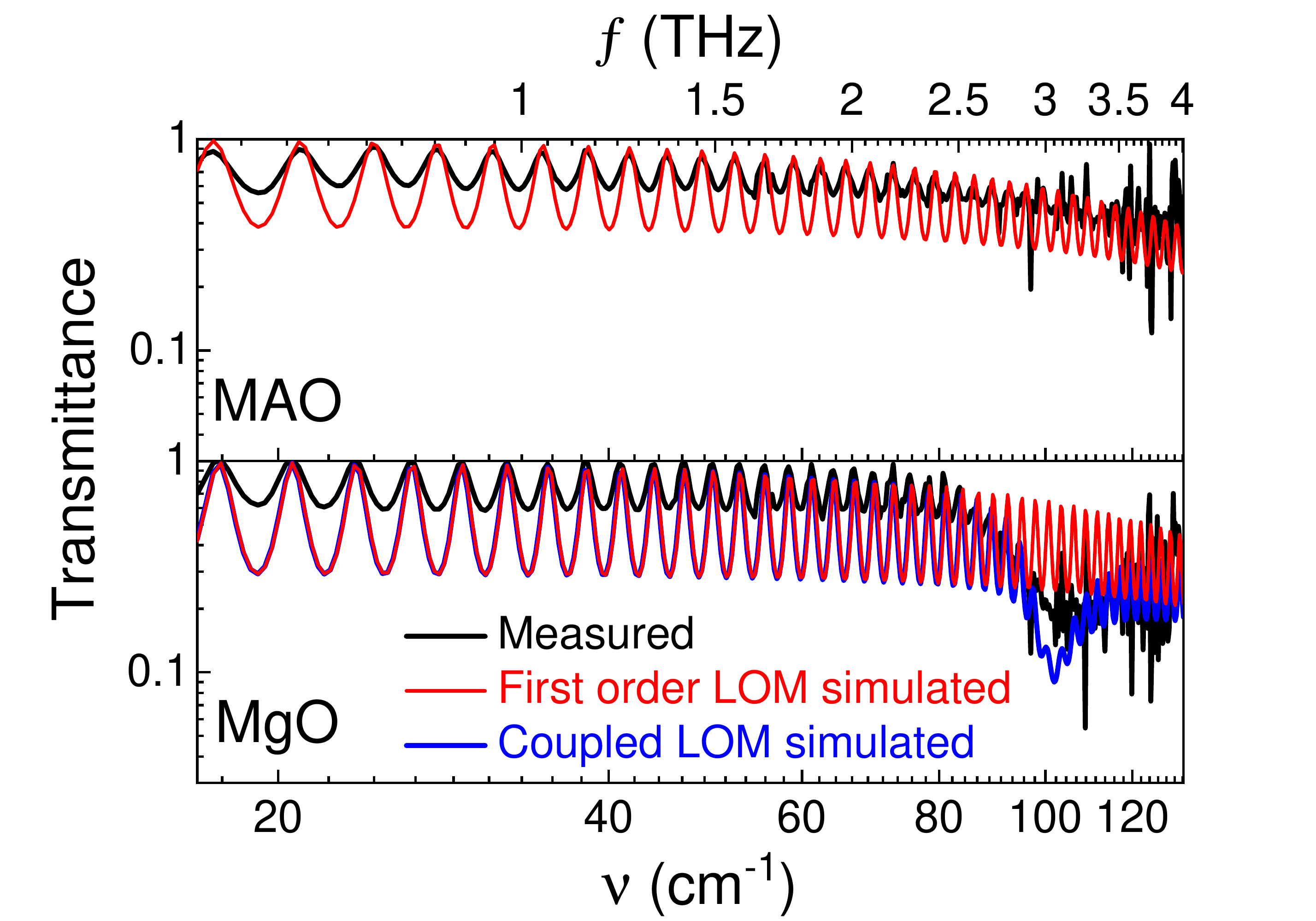}
    \caption{Measured (black) and first order LOM simulated (red) and coupled LOM simulated (blue) transmission spectra in THz regime for MAO (top) and MgO (bottom).}
    \label{fig:my_label}
\end{figure}

\begin{table}[]
    \centering
    \begin{tabular}{c|c|c|c|c|c}
Substrate & Oscillator &    $\Delta \epsilon$ & $\nu$ (cm$^{-1}$) & $\gamma$ (cm$^{-1}$) & $\delta$ \\
\hline
\hline
MAO & TO   & 8.2 & 520 & 10 & -\\
\hline
& TO   & 8.55 & 396 & 4.4 & -\\
& $D_{11}$ & 0.00516 & 97  & 21.15  & -32.5 \\
MgO & $D_{12}$ & 0.0108 & 127  & 64 & - \\
& $D_{21}$ & 0.00221 & 170 & 16.1 & -3.9 \\
& $D_{22}$ & 0.00548 & 193 & 39.15 & -\\
\hline
\hline
    \end{tabular}
    \caption{Parameters obtained from the simulation using Lorentz oscillator model for MAO and MgO.}
    \label{tab:my_label}
\end{table}

In conclusion, the THz emission amplitude from the Pt/Fe/MgO is found to be larger and affected by the induced strain in the Fe spin source layer due to the larger lattice mismatch between Fe and MgO as compared to that between Fe and MAO. This is a result of the larger electrical conductivity (larger electron relaxation time) in the THz frequency regime, which in turn is explained by the larger defect density in Fe layer deposited on MgO. However, the THz bandwidth is is smaller for Pt/Fe/MgO due to the phonon absorption edge at 3 THz (100 cm$^{-1}$) for MgO, which has been confirmed by the coupled Lorentz oscillator model. This study provides an insightful pathway to further engineer the STE metallic layers in terms of microscopic properties to make a powerful spin-based terahertz emitter. 

\begin{acknowledgments}
This work is supported by the Swedish Research Council (grant no. 2017–03799, 2018-04918) and Olle Engkvists Stiftelse (grant no. 182–0365). Stefano Bonetti is acknowledged for fruitful discussion.  
\end{acknowledgments}



\nocite{*}
\bibliography{aipsamp}

\end{document}